
\input harvmac
\hfuzz=100pt
 
\overfullrule=0pt
\parindent=0pt


\def\G(#1){\Gamma(#1)}

\def\C|#1{{\cal #1}}
\def\(#1#2){(\zeta_#1\cdot\zeta_#2)}
\def\lr{\lref}


\def\xxx#1 {{hep-ph/#1}}
\def\lr { \lref}
\def\npb#1(#2)#3 { Nucl. Phys. {\bf B#1} (#2) #3 }
\def\rep#1(#2)#3 { Phys. Rept.{\bf #1} (#2) #3 }
\def\plb#1(#2)#3{Phys. Lett. {\bf #1B} (#2) #3}
\def\prl#1(#2)#3{Phys. Rev. Lett.{\bf #1} (#2) #3}
\def\physrev#1(#2)#3{Phys. Rev. {\bf D#1} (#2) #3}
\def\ap#1(#2)#3{Ann. Phys. {\bf #1} (#2) #3}
\def\rmp#1(#2)#3{Rev. Mod. Phys. {\bf #1} (#2) #3}
\def\cmp#1(#2)#3{Comm. Math. Phys. {\bf #1} (#2) #3}
\def\mpl#1(#2)#3{Mod. Phys. Lett. {\bf #1} (#2) #3}
\def\ijmp#1(#2)#3{Int. J. Mod. Phys. {\bf A#1} (#2) #3}

\def\lam16{\lambda^{16}}

\parindent 25pt
\overfullrule=0pt
\tolerance=10000

\sequentialequations

\noblackbox
\baselineskip 14pt plus 2pt minus 2pt

\Title{\vbox{\baselineskip12pt
\hbox{hep-ph/9807415}
}}
 {\centerline{Unification with  Low  String Scale}}
\bigskip
\centerline{ Constantin P. Bachas }
\medskip
\centerline{Centre de Physique Th{\'e}orique,  Ecole
Polytechnique,}
\centerline{ 91128 Palaiseau, France}
\centerline{\it bachas@pth.polytechnique.fr}
\bigskip
\bigskip

\centerline{\bf Abstract}
I argue that in open-string theory with hierarchically  small (or
large)  extra dimensions, gauge groups can  unify naturally 
with  logarithmically-running  coupling constants at the
high Kaluza-Klein (or string-winding) scale. 
 This opens up the 
possibility of rescuing the standard 
logarithmic  unification  at $M_U\sim 10^{15-18}$ GeV  even if the
fundamental-string scale is much lower, at intermediate or
possibly even electroweak scales.  I also explain, however, why  a low type-I
string scale may  not suffice to obliterate the ultraviolet problems
usually associated with the gauge hierarchy. 

  \vfill\eject


\lr\dien{for a review see
K. Dienes, Phys. Rep. {\bf 287} (1997) 447, hep-th/9602045,
and references therein.}
\lr\koun{S. Ferrara, C. Kounnas and M. Porrati, Phys. Lett. {\bf B206}(1988)
25~; C. Kounnas and M. Porrati, Nucl. Phys. {\bf B310} 355~; 
S. Ferrara, C. Kounnas, M. Porrati and F. Zwirner,
Nucl. Phys. {\bf B318} (1989) 75.}
\lr\witt{E. Witten, Nucl. Phys. {\bf B471} (1996) 135, hep-th/9602070.}
\lr\lyk{J. Lykken, 
Phys.Rev. {\bf D54} (1996) 3693, hep-th/9603133.}
\lr\cb{C. Bachas (1995), unpublished.} 
\lr\ig{I. Antoniadis, Phys. Lett. {\bf B246} (1990) 377.}
\lr\ddg{ K. Dienes, E. Dudas and T. Gherghetta,
hep-ph/9806292~; hep-ph/9803466.}
\lr\bfa{C. Bachas and C. Fabre,
 Nucl.Phys. {\bf B476} (1996) 418, hep-th/9605028.} 
\lr\gp{
M. Bianchi and A. Sagnotti,  Nucl. Phys. {\bf B361} (1991) 519~;  
 Phys. Lett. {\bf
B247} (1990) 517~;  E. Gimon and J. Polchinski,
 Phys. Rev. {\bf D54}  (1996) 1667, hep-th/9606176.
}
\lr\aadd{I. Antoniadis, N. Arkani-Hamed, S. Dimopoulos and  G. Dvali,
hep-ph/9804398.}  
\lr\add{I. Antoniadis and K. Benakli,
Phys. Lett. {\bf B326}  (1994) 69, 
 hep-th/9310151~;  I. Antoniadis, K.
    Benakli and M. Quir{\'o}s,  Phys. Lett. {\bf B331} (1994) 313, 
 hep-ph/9403290~; 
I. Antoniadis, S. Dimopoulos and  G. Dvali, Nucl.Phys. {\bf B516}  (1998) 70, 
hep-ph/9710204, N. Arkani-Hamed, S. Dimopoulos and  G. Dvali,
hep-ph/9807344.}  
\lr\ark{N. Arkani-Hamed, S. Dimopoulos and  G. Dvali,
hep-ph/9803315.}
\lr\tye{G. Shiu and  S.-H.H. Tye,
hep-th/9805157.}
\lr\kaplu{V. Kaplunovsky,  Nucl. Phys. {\bf B307} (1988)
 145~; {\it ibid} {\bf
B382} (1992) 436 (E).}
\lr\gin{  P. Ginsparg, Phys. Lett. {\bf B197} (1987) 139.  }
\lr\ablt{I. Antoniadis, C. Bachas, D. Lewellen and T. Tomaras,
 Phys. Lett. {\bf B207} (1988)
 441~; C. Bachas,
in Proceedings of the 9th Workshop on Grand
Unification, Aix-Les-Bains 1988,  World Scientific.
}
\lr\dl{M. Douglas and M. Li,   hep-th/9604041~; 
C. Bachas and  E. Kiritsis, 
Nucl. Phys. Proc. Suppl. {\bf 55B} (1997) 194-199, hep-th/9611205.}
\lr\bfy{C. Bachas, C. Fabre and T. Yanagida,
Phys. Lett. {\bf B370} (1996) 49, hep-th/9510094.} 
\lr\chir{C. Angelantonj, M. Bianchi, G. Pradisi, A. Sagnotti and
Ya. S. Stanev, Phys. Lett. {\bf B385} (1996) 96, hep-th/9606169~;
Z. Kakushadze and  G.  Shiu, Phys. Rev. {\bf D56}  (1997) 3686,
hep-th/9705163, and hep-th/9706051~;
G. Zwart, hep-th/9708040~; 
G. Aldazabal, A. Font, L. E. Ibanez and  G. Violero, hep-th/9804026~;
 Z. Kakushadze, G.  Shiu and
S.-H.H. Tye,  hep-th/9804092~; Z. Kakushadze, hep-th/9804110~.}
\lr\efn{ J. Ellis, A. Faraggi and D. Nanopoulos, Phys. Lett. {\bf B419}
(1998) 123, hep-th/9709049.}
\lr\kap{ E. Caceres, V. Kaplunovsky and M. Mandelberg,
Nucl. Phys. {\bf B493}(1997) 73, hep-th/9606036.}
\lr\kir{C. Bachas and E. Kiritsis,
Nucl.Phys.Proc.Suppl. {\bf 55B} (1997) 194, hep-th/9611205.}
\lr\abfpt{ I. Antoniadis, C. Bachas, C. Fabre, H. Partouche and
T. Taylor, Nucl.Phys. {\bf B489}  (1997) 160, hep-th/9608012.}
\lr\dlk{ L. Dixon, V. Kaplunovsky and J. Louis, Nucl. Phys. {\bf
B355}(1991) 649.}


\noblackbox
\baselineskip 14pt plus 2pt minus 2pt

The perturbative
unification of the standard-model gauge coupling constants 
 extrapolated to  $M_U \sim
10^{15-18}$ GeV ,  is one of the  few solid hints  we have about physics
beyond the standard model. The weakly-coupled heterotic string,
compactified near the string scale $M_{s}^{het} \sim 10^{17-18}$ GeV,
is naturally compatible with such a unification
scenario \kaplu\gin  -- the  one-to-two order-of-magnitude
discrepancy  between $M_U$ and $M_{s}^{het}$
 being a relatively minor  nuisance when
compared to the logarithmic distance separating  them
from the electroweak scale \dien.
More recently there has been renewed interest in the possibility
of departing from this traditional point of view, and allowing
hierarchically-different  string and compactification scales
\witt\lyk\kap\cb\ark\aadd\tye. It has been argued in particular that the
string scale could  be lowered in a controlled way, either 
 in type-I theory \witt\lyk\aadd\tye~
where this is  allowed by the tree-level relations, 
 or in the heterotic theory \cb~  where  the tree-level
 relation   ($M_{s}^{het}\sim g_{\rm YM}
M_{\rm Planck}$)   seems to forbid  it. In the latter case 
 one must   take  hierarchically-small bare Yang-Mills couplings
 and then invoke large-radius
 threshold effects to drive some of them
to their  observed  low-energy values. 
The fact that such  exotic
scenaria are not immediately ruled out by experiment  
and might be even testable   in the near future  \add,
is certainly sufficient  motivation by itself to continue 
investigating  them further.

 These   scenaria share  with the more conservative
point of view  the usual problem  of  vacuum
stability after  supersymmetry breaking. 
They seem to face furthermore  some   extra difficulties~:
if one  abandons, in particular,   the conventional
unification at high $M_U$
by logarithmic running, one gives up a `natural' explanation for the value of 
${\rm sin}^2\theta$, while
 facing at the same time   the risk of
totally uncontrollable  proton decay. 
Ideas on how to proceed have been discussed  in references
 \ddg\tye\aadd\efn.
It has been  argued in particular   that large-radius threshold effects
can  accelerate unification  without affecting  the value of 
${\rm sin}^2\theta$ \ddg, or  that this latter may be fixed by
tree-level string relations without group unification \tye. 
Proton decay,  on the other hand, could be
conceivably suppressed  by discrete gauge symmetries \aadd,
by a weakly-gauged  custodial U(1) \efn\tye, or by the
higher-dimensional  nature of the  interactions \ddg.

 One assumption  often  made in these  discussions
is  that the string scale sets the
 ultraviolet cutoff, and that logarithmic  unification can be
 at best achieved at or below this  scale. Here  I
 will try to explain why  in (weakly-coupled) type-I theory this naive
 intuition need not be valid.
 I will show  in  particular, using the results of \bfa,
 that  in the presence of hierarchically small or large extra dimensions, 
some 4d  gauge couplings can   
run logarithmically and unify at the super-heavy  Kaluza-Klein, respectively
 string-winding  scale. This opens up the possibility of rescuing
in a  modified form
the successful unification at  $M_U\sim 10^{15-18}$ GeV,   even 
if the type-I string scale ($M_s^I$)  is much lower. in the
intermediate or even in the TeV range. More generally, however, I will
show that 
 superheavy ($M\gg M_s^I$) momentum or winding
modes can  give
power-like radiative corrections to  4d gauge couplings -- a warning that
lowering the type-I string scale does   not  suffice to obliterate the
UV  problems, which  are  usually associated with 
the existence of the gauge hierarchy.

Let me first explain the main  point  in the simplest  context of a
 $T^4/Z_2\times T^2$ type-I
 compactification, with N=2 supersymmetries in four dimensions \gp.
These are the models whose threshold effects have been  analyzed in
 ref. \bfa.
The background has 32 D9-branes, as well as 32 D5-branes transverse to
 the $T^4/Z_2$ orbifold. Let $g_s$ be the string coupling constant, and
  $v_4$ and $v_2$ be, respectively,
 the volume of the orbifold and the area of the
 two-torus, both measured in units of 
$2\pi\sqrt{\alpha^\prime}$ (from now on and
unless otherwise stated, $\alpha^\prime$
 will stand for the type-I string scale).
Dualizing the two coordinates of the torus  gives an equivalent type-I'
 background with D7-branes and D3-branes, and with
\eqn\T{
{\tilde v_2} = 1/v_2  \ , \ \ {\rm and}\ \ \ 
{\tilde g_s}  = g_s/v_2 \ .
}
There are two types of gauge groups -- those living on the D5-branes
or equivalently the type-I' D3-branes, 
 and those  living on the D9-branes or type-I' D7-branes. 
The Planck mass  and the two
tree-level Yang-Mills couplings in four dimensions read 
\eqn\couplings{
g^2_{\rm YM}\Bigl\vert_{D3}  = 
 v_4\; g^{ 2}_{\rm YM}\Bigr\vert_{D7}  =    2\sqrt{2} \pi\;
 {\tilde g_s}  \ , \ \ \ \ 
 {\rm and}\ \ \ 
M^2_{\rm Planck} =
 {v_4 {\tilde v_2} \over 2\pi\alpha^\prime {\tilde g_s}^2} \ .
}
Taking  ${\tilde v_2}\gg 1$,  with $v_4$ and   ${\tilde g_s}$ 
of  order one,  creates a hierarchy between  the 4d Planck scale and 
the string scale.  The Kaluza-Klein excitations 
 on $T^2$  (which become winding states  in the  type-I' picture)
have masses   
$M_{\rm KK}  \sim \sqrt{{\tilde v_2}/\alpha^\prime}$
and are thus super-heavy in this limit.
The key remark for our purposes here is that  the logarithmic
running of the four-dimensional 
gauge couplings does not stop at the (much lower) string scale,  but
may continue  all the way up to  $M_{\rm KK}$.

  To understand why notice first  that, because of  N=2
supersymmetry, only short BPS states  renormalize the gauge
coupling constants.
Since all excited states of an open string are
non-BPS,  the fundamental
string scale drops  out completely from the loop  corrections \dl\bfa.
The final expression at one-loop is formally identical 
to that of a six-dimensional gauge field theory, compactified down to four
dimensions on a (tiny in type-I language)
 two-torus, and with the  quadratic divergence subtracted out.  
To be more specific, take the torus to be
orthogonal and let  $R_1$ and $R_2$ be its two radii  (so that $v_2 =
R_1R_2/\alpha^\prime$). The one-loop contribution of a
mutliplet in the $r$th representation of some  gauge-group factor reads
\bfa
\eqn\form{
\Delta {4\pi^2\over g^2_{YM}(\mu)}
 = 
\pm  C_r \int_0^\infty {dt\over t}\;  e^{-(\mu^2 +M_r^2)  t}\;
\left( \sum_{n_1,n_2\in Z}  e^{- \left[ ({n_1\over R_1})^2 +
 ({n_2\over R_2})^2 \right]  t }\ 
- { \pi R_1R_2 \over t} \right)\ ,
}
where $C_r$  is the quadratic Casimir, $M_r$ is the mass of the 
mutliplet, $\mu$ is an infrared cutoff,
and the sign is plus  for hypermutliplets and minus for the
(adjoint)  vector. The ${1/ t}$ subtraction terms, transformed to the
transverse closed-string channel, 
sum up over all representations to zero
 -- this is guaranteed by 
 tadpole cancellation
as  explained in ref. \bfa. 
For inverse radii roughly equal, and much larger  than the  other mass 
scales,  we may think of  the
expression in parenthesis as a particular way of imposing an
ultraviolet  (Pauli-Villars type) 
cutoff $M_{KK}\sim 1/R_i$ to the standard  four-dimensional 
loop corrections.  The gauge-coupling constants
will thus  run  up to this 
cutoff, much higher than string scale, as advertized.

If the two
radii are hierarchically different, say $R_1\gg R_2$, then the
lighter Kaluza-Klein modes (with masses $n_1/R_1$) will 
contribute effectively  up to the heavier, $1/R_2$,  scale. 
The theory is in this range five-dimensional  and the couplings
`evolve' with a power-law, i.e.  receive  large (non-logarithmic)  threshold
corrections. This is the scenario of  accelerated unification
considered  in refs. \ddg  -- but with  the role of the UV cutoff
played again by the heaviest  Kaluza-Klein scale,  rather than by the
tension of the type-I string. These huge power-law corrections will
survive even in the limit in which {\it all} Kaluza-Klein states are
sent to infinity --  a warning
that the gauge hierarchy may not be nullified  simply by lowering the
type-I string scale. In order to have only  log-corrections, we
must keep  the (complex-structure) modulus
$R_2/R_1$  of order one -- i.e. insist that the only runaway
modulus be  $v_2$.

   The above  results can be understood heuristically as follows:
for generic type-I  Wilson lines  (i.e. for generic separations
of the type-I'  D-branes on the torus)
 the mass of the charged gauge bosons is of the order of the
  Kaluza-Klein scale.
It is therefore not  surprising that the gauge couplings  unify in
general   up there, and  not at the much lower fundamental
string scale. Furthermore, 
 the ultraviolet regime of open-string
theory is governed by the infrared regime of (super)gravity. Turning
on a background gauge-field strength perturbs the equilibrium of the
D-branes, which in the type-I' language live in a large transverse
two-dimensional space. The leading ($F^2$)  perturbation to  vacuum
energy exhibits therefore
the  characteristic two-dimensional  logarithmic dependence, and can
even grow linearly with radius if the torus becomes effectively
one-dimensional at larger scales. Note that in  applying  the argument to
 D3 branes one must recall that they couple to
twisted  massless closed-string
modes, which  are localized at  orbifold fixed points. Thus even if
the orbifold  is  also large, the dominant effect stays  logarithmic.

Such  heuristic considerations can  be perhaps  extended to stable
N=1 or even N=0 
backgrounds. A  more precise  argument  can be given for 
N=1 orbifold vacua at one-loop order: in type-I' language these vacua
are made out of  D3-branes, D7-branes transverse to
one  of the three complex  compactified dimensions, and
corresponding orientifolds.
The one-loop open-string diagrams  can be classified according to the amount
of supersymmetry they preserve. For instance, a (33)  amplitude
without an orbifold   twist  preserves N=4 supersymmetry,
while a
(37) amplitude without twist  or a (33) amplitude with a $Z_2$ twist
preserve N=2 supersymmetry. 
Now the contributions of N=4 sectors vanish because of helicity-supertrace
formulae \kir, while those of N=1 sectors do not depend 
 on the compactification
moduli. The N=2 contributions, on the other hand, are blind to the excited
open-string states  and will 
be  cut off at high energy by the appropriate 
Kaluza-Klein scale, as already explained. Thus, if all
complex-structure moduli are order one, the gauge couplings should run
logarithmically, first with the N=1 $\beta$-functions up to string
scale, and then with modified $N=2$ $\beta$-functions up to the
much heavier  Kaluza-Klein scale. This should be straightfoward to
check in some of the explicit N=1 type I models \chir.

   In  light of the heterotic/type-I duality, one can ask  how
   this picture would look from the heterotic side. For the
$T^4/Z_2\times T^2$  compactifications, the dictionnary between
   various moduli is as follows \abfpt~:
\eqn\dual{ S_I = S_{het}\, \ \ \ 
  S^\prime_I = T_{het}\, \ \ {\rm and} \ \ \ 
  U_I = U_{het} \ ,
}
where the real parts of the $S$'s  determine as usual the 4d Yang Mills
   couplings (there are two of those on the type I side, corresponding
   to D9-brane  and D5-brane gauge groups), the $U$'s are
   complex-structure moduli, and $T_{het}$ is the torus-area modulus
   measured in units of the heterotic string scale. Assuming
   the  type-I gauge couplings  to be of order one  implies that
   $T_{het}$ is also order one, i.e. that the two-torus is roughly of
   heterotic-string size.  In the limit of interest ($v_2\ll 1$)
   this is  hierarchically smaller than  the  type-I string size,
   which is in turn comparable to the orbifold volume ($v_4\sim 1$). 
   To keep $S_{het}$ of order one we must, furthermore,  take the heterotic string
  coupling  hierarchically large. 
   We are thus dealing with a very strongly-coupled heterotic string vacuum,
   compactified on a very large $Z_2$ orbifold -- a situation in which
  the semiclassical approximation  cannot  be trusted. Nevertheless,
  thanks to the extended N=2 supersymmetry, the one-loop corrections
  to gauge couplings are (perturbatively) exact, and in this case are furthermore
   independent of the orbifold volume \dlk.
   They  evolve, therefore,  logarithmically up to the heterotic string
   cutoff,  $M_s^{het}$, which is of the same order as the 
   Kaluza-Klein scale  on the two-torus. This is precisely what we
  found on the type-I side.

   The fact that 
   one-loop threshold effects can be   insensitive to
    certain moduli  was used in ref. \ig~ in an early effort to construct
     heterotic vacua that  are perturbatively controllable, 
    despite a  large compactification scale. As our discussion here
   shows, this may work for quantities protected by  extended supersymmetry,
   but is problematic otherwise. We have indeed argued that the
   (light) open-string  scale should enter  as an ultraviolet cutoff
   in N=1 sectors of an orbifold, an effect that is invisible on the
   heterotic side since  light open
   strings cannot be  captured by the heterotic loop expansion,

Let me conclude with some  comments  on how the unification scenario  may apply to 
 a realistic type-I string model. Clearly, in view of the above
 discussion, the gauge couplings can at best  run logarithmically with the
 MSSM $\beta$-functions up to the type-I string scale. They could  then
 continue their logarithmic running  up to the heavier unification scale, but  with 
 modified  $\beta$-functions coming from certain N=2 sectors of the
 theory -- those sensitive to the large Kaluza-Klein scale. Now the low-energy
 prediction for ${\rm sin}^2\theta$ depends only
on the ratio $(\beta_1 -\beta_2)/(\beta_2-\beta_3)$ -- a quantity that is (numerically)
 robust under  the addition of complete SU(5)
 representations, or under minimal N=2 extensions  of the MSSM  particle
 content \ddg, see also  \bfy. It is thus  plausible to  hope that the 
successful prediction of ${\rm
 sin}^2\theta$ can be retained despite the $\beta$-function
 modification at $M_s^I$.  Pushing the unification scale up  can 
 also suppress the conventional channel of  proton decay,  through
 exchange of heavy (X and Y) gauge bosons. Since the type I scale
 is however low, one would still need to suppress the dangerous channels
 mediated by exchange of excited open strings.

\vskip 0.5cm

 {\it Aknowledgements}:
 I  thank K. Dienes  for  a discussion,  I. Antoniadis and  
 J. Ellis for remarks  on an  earlier version  of the manuscript, and
 particularly K. Benakli for correcting a premature  claim about
 proton decay.  This work was partially supported by
the EEC grant  TMR-ERBFMRXCT96-0090.

\listrefs

 \bye